\documentclass[sigchi]{acmart}
\AtBeginDocument{%
  }

\setcopyright{acmlicensed}
\copyrightyear{2025}
\acmYear{2025}
\acmDOI{XXXXXXX.XXXXXXX}
\acmISBN{978-1-4503-XXXX-X/2018/06}

\usepackage{booktabs}
\usepackage{tabularx}
\usepackage{subcaption}
\usepackage{multirow}
\usepackage{array}
\newcolumntype{Y}{>{\raggedright\arraybackslash}X}

\begin{document}

\title{Object-Driven Narrative in AR: A Scenario-Metaphor Framework with VLM Integration}

\author{Yusi Sun}
\orcid{1234-5678-9012}
\affiliation{%
  \institution{the University of Hong Kong}
  \city{Hong Kong}
  \country{China}
}
\email{soniasun@connect.hku.hk}

\author{Haoyan Guan}
\affiliation{%
  \institution{the University of Hong Kong}
  \city{Hong Kong}
  \country{China}
}
\email{hyguan@hku.hk}

\author{leith K.Y. Chan}
\affiliation{%
  \institution{the University of Hong Kong}
  \city{Hong Kong}
  \country{China}
}
\email{lkychan@hku.hk}

\author{Yong Hong Kuo}

\affiliation{%
  \institution{the University of Hong Kong}
  \city{Hong Kong}
  \country{China}
}
\email{yhkuo@hku.hk}


\begin{abstract}
Most adaptive AR storytelling systems define environmental semantics using simple object labels and spatial coordinates, limiting narratives to rigid, pre-defined logic. This oversimplification overlooks the contextual significance of object relationships-for example, a wedding ring on a nightstand might suggest marital conflict, yet is treated as just "two objects" in space. To address this, we explored integrating Vision Language Models (VLMs) into AR pipelines. However, several challenges emerged: First, stories generated with simple prompt guidance lacked narrative depth and spatial usage. Second, spatial semantics were underutilized, failing to support meaningful storytelling. Third, pre-generated scripts struggled to align with AR Foundation's object naming and coordinate systems. We propose a scene-driven AR storytelling framework that reimagines environments as active narrative agents, built on three innovations: 1. State-aware object semantics: We decompose object meaning into physical, functional, and metaphorical layers, allowing VLMs to distinguish subtle narrative cues between similar objects. 2. Structured narrative interface: A bidirectional JSON layer maps VLM-generated metaphors to AR anchors, maintaining spatial and semantic coherence. 3. STAM evaluation framework: A three-part experimental design evaluates narrative quality, highlighting both strengths and limitations of VLM-AR integration. Our findings show that the system can generate stories from the environment itself, not just place them on top of it. In user studies, 70\% of participants reported seeing real-world objects differently when narratives were grounded in environmental symbolism. By merging VLMs' generative creativity with AR's spatial precision, this framework introduces a novel object-driven storytelling paradigm, transforming passive spaces into active narrative landscapes.
\end{abstract}


\begin{teaserfigure}
  \includegraphics[width=0.9\textwidth]{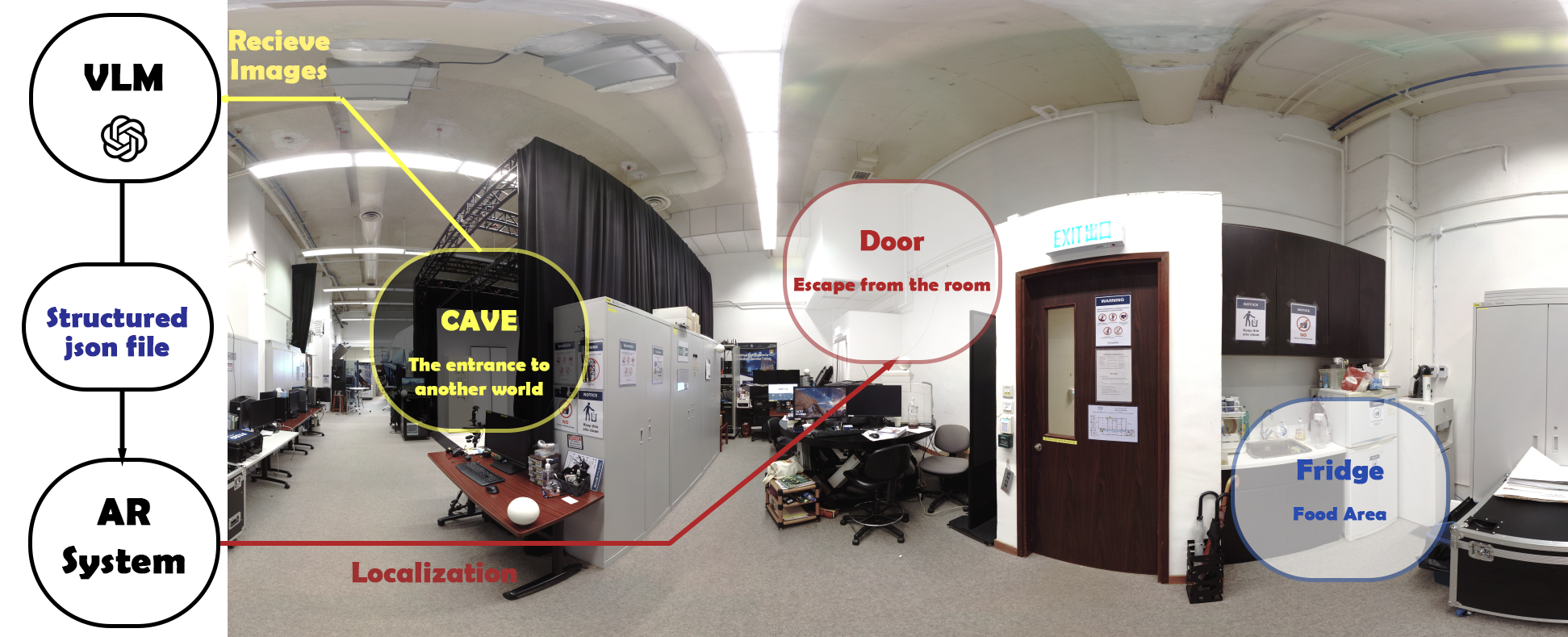}
  \caption{The overall structure of our proposed pipeline. By discovering unique objects in the environment and their metaphors, the VLM is able to create relevant storylines and send them to the AR system to generate the corresponding prefabs}
  \Description{It's a simply introduce about our pipeline.}
  \label{fig:teaser}
\end{teaserfigure}


\maketitle

\section{Introduction}

As a medium of spatial narrative, augmented reality (AR) has demonstrated its unique value in the fields of education, cultural heritage, tourism, psychotherapy and so on \cite{cassell2001making, Grasset2008, Billinghurst2001}. In early childhood education, immersive storytelling experiences supported by AR have been shown to be effective in promoting the development of imagination \cite{cassell2001making}. By enabling the alignment, integration, and fusion of virtual narrative spaces with physical environments, AR opens up new possibilities for experiential storytelling.

Previous research has extensively focused on how AR content can be dynamically adapted based on player behavior and environmental context using semantic information. While current systems—such as the early work by Wanwan Li et al. \cite{liwanwan2023}, demonstrate the ability to adjust AR content through semantic perception, their modeling of spatial semantics remains superficial. Typically limited to object-level labels (e.g., “chair”), these approaches neglect the narrative relational space—that is, the human cognitive associations triggered by the implicit configuration of objects in a scene. This limitation leads to two major problems: 1. loss of scene sensitivity: Narrative intent for objects of the same category becomes conflated across contexts (e.g., a kitchen knife and a hidden bedroom knife are both labeled simply as “knife”). 2. Suppress emergent storytelling: Metaphorical juxtapositions fail to elicit deeper narrative meaning (e.g., a wedding ring and a bottle of pills on a nightstand do not trigger associations with crisis or loss).

With the emergence of VLMs such as GPT, AI has gained enhanced capacity for spatial understanding and contextual narrative generation, enabling new paradigms in story construction. However, integrating VLMs into AR introduces specific technical challenges: When narratives are pre-generated, VLMs cannot accurately align with AR Foundation’s runtime object names, 3D positions, or interaction mappings. But in real-time generation, it's difficult to ensure the coherence and completeness of the evolving storyboards.

These challenges raise two key challenges:
\begin{itemize}
    \item \textbf{Challenge 1:} How to disambiguate narratives among visually similar objects using metaphorical reasoning?
    \item \textbf{Challenge 2:} How to preserve VLM creativity while ensuring physical plausibility within AR environments?
    \item \textbf{Challenge 3:} How to bridge a generative and structured system between VLM and Unity?
\end{itemize}

To address these challenges, this study proposed a scenario-metaphor-driven AR story generation framework. This system adapted VLM outputs by incorporating name, interaction, and location modules in a structured AR compatible format. As shown in Figure \ref{fig:workflow}, our approach takes spatial images or video as input. The VLM first identifies key metaphorical objects, then generates a spatial narrative informed by both visual context and implied meaning.

Crucially, the metaphorical layer allows the system to distinguish context-specific narrative roles for visually similar objects. To ensure spatial coherence, this study designed a structured JSON file that maps metaphorical output to AR constraints (e.g., spatial anchoring, occlusion management), enabling end-to-end generation with real-time AR deployment. In user tests, this method achieved a spatial consistency score of 5.31/7, indicating strong alignment between virtual narratives and real-world space.

\begin{figure}[t]
  \centering
  \includegraphics[width=\linewidth]{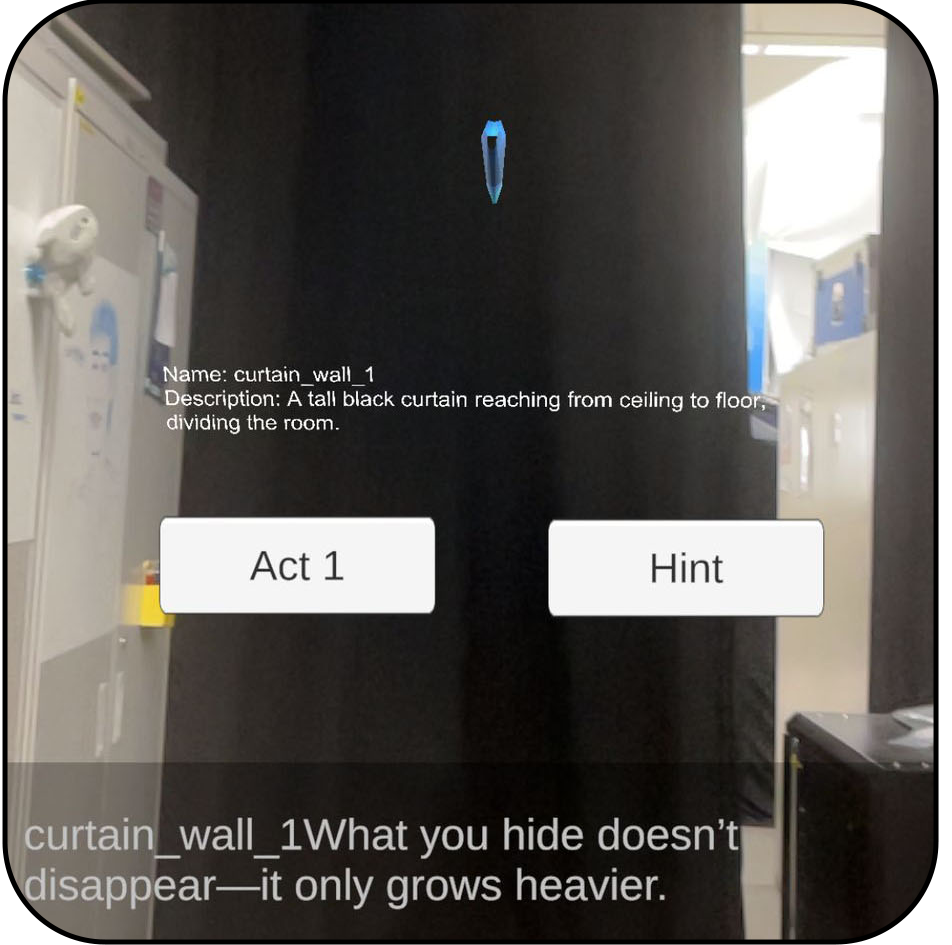}
  \caption{The example Scenario of System Integration Validation}
  \Description{1 key object in System Integration Validation and its description}
  \label{fig:test3}
\end{figure}

The effectiveness of our framework is validated through three designed experimental modules: basic VLM capability assessment (STAM metric), AR storyboard assessment, and real-world AR deployment (shown in \ref{fig:test3}). Therefore, the main contributions of this paper include:

\begin{enumerate}
\item This paper posed a new research question: How can XR narratives be enhanced by leveraging natural narrative relationships between real-world objects?

\item This paper developed and optimize an AR story generation pipeline, integrating VLMs while addressing the challenges of metaphor-driven narratives and spatial realism.

\item This paper introduced a metaphorical object layer, enabling VLMs to differentiate narrative roles among objects with similar surface properties, thereby generating scripts suited for AR storytelling.

\item This paper validated our approach through three complementary experiments, providing empirical insights into both spatial understanding and narrative generation using VLMs.
\end{enumerate}

\section{Related Work}
\subsection{Semantically understanding in AR}

Semantic-based AR technologies typically leverage the semantic and spatial attributes of objects, using semantic understanding techniques to present AR content. Early research on semantic understanding in AR focused on object-level attribute mapping, often using explicit rules to achieve basic environmental adaptation. For example, Han et al. \cite{han2020live} simulated realistic AR content trajectories by assigning different material properties to visually engaging physical objects. Retargetable AR \cite{tahara2020retargetable} adjusted the behavior of AR virtual characters based on the functions of nearby objects. Researchers also implemented optimization algorithms to enhance adaptability.

As semantic understanding technologies advanced, the focus shifted toward environment-level semantic associations. Liang et al. \cite{liang2021scene} developed an algorithm to synthesize AR pet behavior based on indoor layouts. Lang et al. \cite{lang2019} and SemanticAdapt \cite{cheng2021} considered the semantic and functional relationships of physical objects to optimally place MR assets and interfaces around the user. Lindlbauer et al. \cite{Lindlbauer2019} adjusted the level of detail in MR interfaces based on the semantic implications of human activity. Tailor Reality \cite{Dong2021} further used human visual perception of different entities as semantic cues to reconfigure physical layouts within MR environments. Qian et al. \cite{qian2022} introduced a temporal creative process to eliminate implicit semantic ambiguities, extending semantic-level AR content associations to the entire environment. Tahara et al. \cite{Tahara2020} used 3D scene graphs to link AR content with the physical world for more natural spatial arrangements. Changyang Li et al. \cite{Li2022} considered the spatiotemporal relationships of abstract narrative event sequences to support a variety of activity possibilities. Wanwan Li et al. \cite{liwanwan2023} explores how to adapt narratives to real-world locations via an optimization. Ziming Li et al. \cite{Li2025} explored how LLM-powered agents can perform environment-aware spatial interactions and conversations.

These AR experiences link the functions of AR content to the functions of surrounding physical objects, enabling accurate deployment across diverse environments. However, in most semantic-based systems, adaptation remains limited to object-level semantic associations. Most approaches reduce semantics to object labels (e.g., “chair,” “table”), lacking the ability to distinguish narrative differences between objects of the same category (e.g., a kitchen knife vs. a bedroom knife). This often leads to a lack of scene sensitivity and deeper narrative expression. Furthermore, many methods rely on predefined rules \cite{Tahara2020, Li2025}, making it difficult to capture implicit semantic relationships in dynamic environments. However, as our knowledge, no work has addressed how to leverage deep semantic relations, between objects in natural environments for AR storytelling purposes. Hong et al. \cite{HongNEURIPS2023} have demonstrated that VLMs possess fine-grained spatial understanding capabilities. Building on this, the innovation of this paper lies in combining deep semantic parsing from VLMs with probabilistic narrative reasoning, thereby overcoming the limitations of traditional label-based semantics.

\subsection{Interactive Narratives}

\subsection{System Implementation}
\begin{figure}[htbp]

  \centering
  \includegraphics[width=\linewidth]{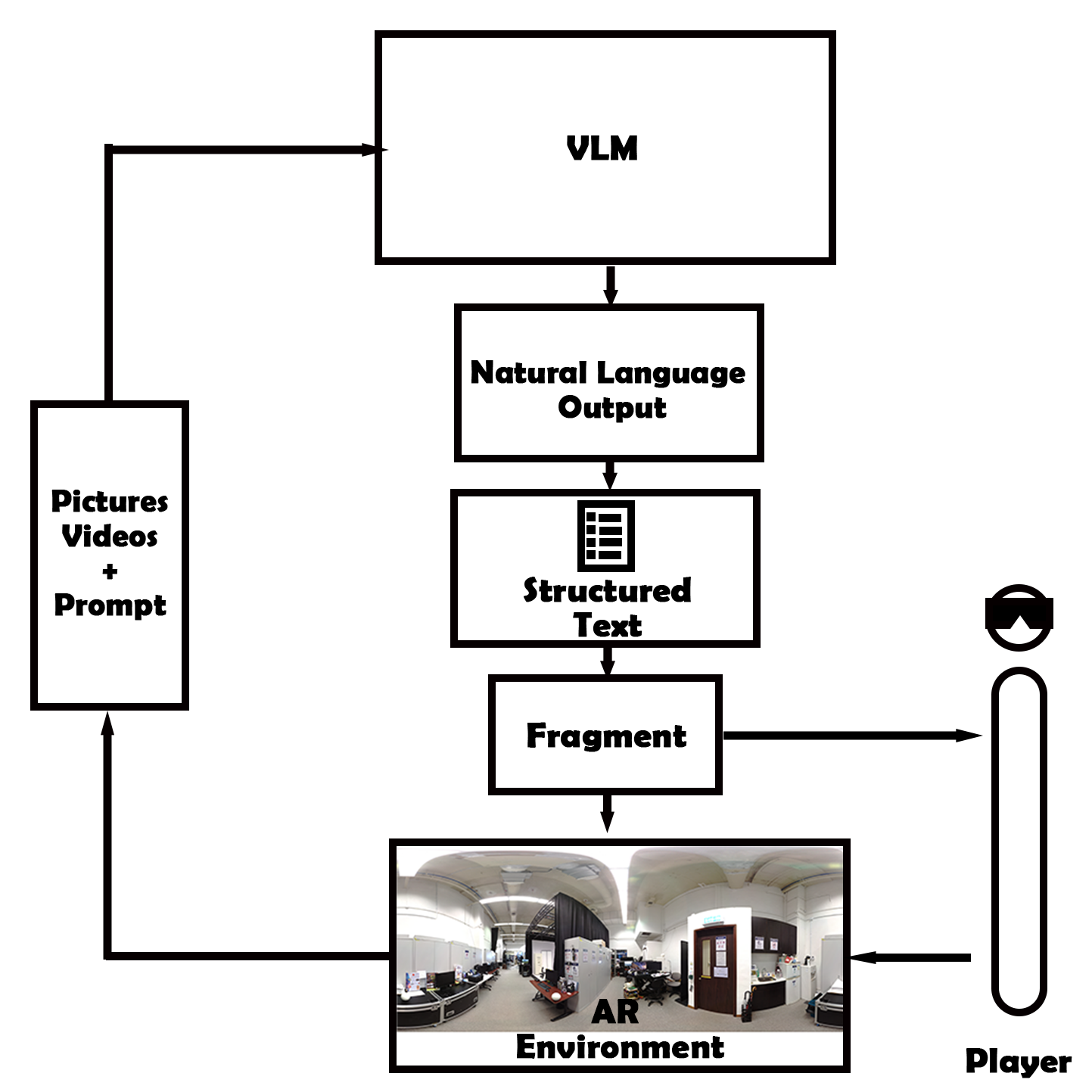}
  \caption{The communication flow between AR application and VLM}
  \Description{It describes how the AR application send message to VLM and get response from it.}
  \label{fig:workflow}
\end{figure}

Interactive narrative, as a core form of digital experience, enables the dynamic evolution of storylines through user interaction with the story world\cite{Riedl_Bulitko_2012}. Its central challenge lies in balancing user agency with narrative coherence. Traditional branching story graph approaches maintain authorial intent through predefined decision points\cite{Riedl_Bulitko_2012}, but face scalability issues due to exponential growth in narrative paths — N choice points can lead to $2^n$ possible storylines — thus prompting the development of automatic story generation techniques. Consequently, the evolution of interactive storytelling consistently revolves around a core question: How can narrative artistic integrity be preserved within open-ended user participation?

Early researchers constructed narrative logic through rule-based systems. For example, HEFTI\cite{Ong2004} used genetic algorithms to recombine story units, while El-Nasr et al.\cite{el2007interaction} designed an interaction framework grounded in dramatic theory. These pioneering efforts laid the foundation for rule-driven generative paradigms. With the rise of mobile AR technologies, spatial context awareness emerged as a key focus. The LARAT tool developed by Ruminski's team\cite{Rumiski2013} was among the first to integrate geographical location into AR storytelling. Chen et al.\cite{chen2021} later implemented scene-based micro-narrative generation in SceneAR. At the same time, Li et al.\cite{Li2022} proposed a real-time narrative adjustment algorithm, while Wanwan Li et al.\cite{liwanwan2023} introduced an adaptive framework for environment-aware storytelling—both signaling a deep integration of optimization techniques into spatial narrative. In recent years, the rise of deep learning and large language models (LLMs) has led to a data-driven paradigm shift. Wang et al.\cite{pengchengwang2017, pengchengwang2018} applied LSTM networks to achieve personalized narrative planning. Meanwhile, Chung’s team\cite{Chung2022} conducted a systematic evaluation, revealing potential limitations of LLMs in maintaining controllability over long-form narratives.

Despite ongoing technological advances, current AR narrative systems face several key limitations: 1. Most methods\cite{Li2022, Rumiski2013} still rely on object-level labels, failing to capture metaphorical semantics embedded in environmental configurations. 2. While some recent work\cite{Han2025} explores multimodal generation using VLMs, it has yet to deeply integrate spatial understanding with narrative reasoning. 3. Existing optimization frameworks\cite{liwanwan2023} tend to focus narrowly on physical spatial alignment, overlooking the narrative emergence triggered by object juxtaposition. These limitations highlight a critical theoretical gap in current systems concerning environmental narrative intelligence, underscoring the urgent need for a paradigm shift to address these shortcomings.

\section{Methodology}

This study first built a pipeline that integrate VLM into traditional AR story generation pipeline. However, several questions raised up:
When narratives are pre-generated, VLMs cannot accurately align with AR Foundation’s runtime object names, 3D positions, or interaction mappings, but in real-time generation, it's difficult to ensure the coherence and completeness of the evolving storyboards. 

As illustrated in Figure \ref{fig:workflow}, this study presented a novel integration of VLMs into the traditional AR content pipeline, addressing the dual challenges of narrative generation and spatial adaptation to solve the challenge 2. Compared to previous work such as Li et al.\cite{Li2022}, which handled spatial sampling and script generation as separate processes, our system enables coordinated optimization through a unified VLM-based approach. By taking scene images as input and combining them with user interactions, the system dynamically generates 3D spatial narratives.

The core innovation lies in the intermediary layer, where this study used a structured JSON file to achieve precise semantic-to-spatial mappings in Figure \ref{fig:workflow}. This ensures accurate AR content placement while preserving VLM creativity.

Structured JSON Architecture
The structured JSON file bridges VLM output and the AR environment by organizing content into three types:

\begin{itemize}
    \item object: key narrative items selected by GPT.
    \item mainstory: a linear narrative generated around those objects.
    \item fragments: additional object descriptions generated to support theme comprehension and player engagement.
\end{itemize}

\begin{figure}[tbp]
  \centering
  \includegraphics[width=\linewidth]{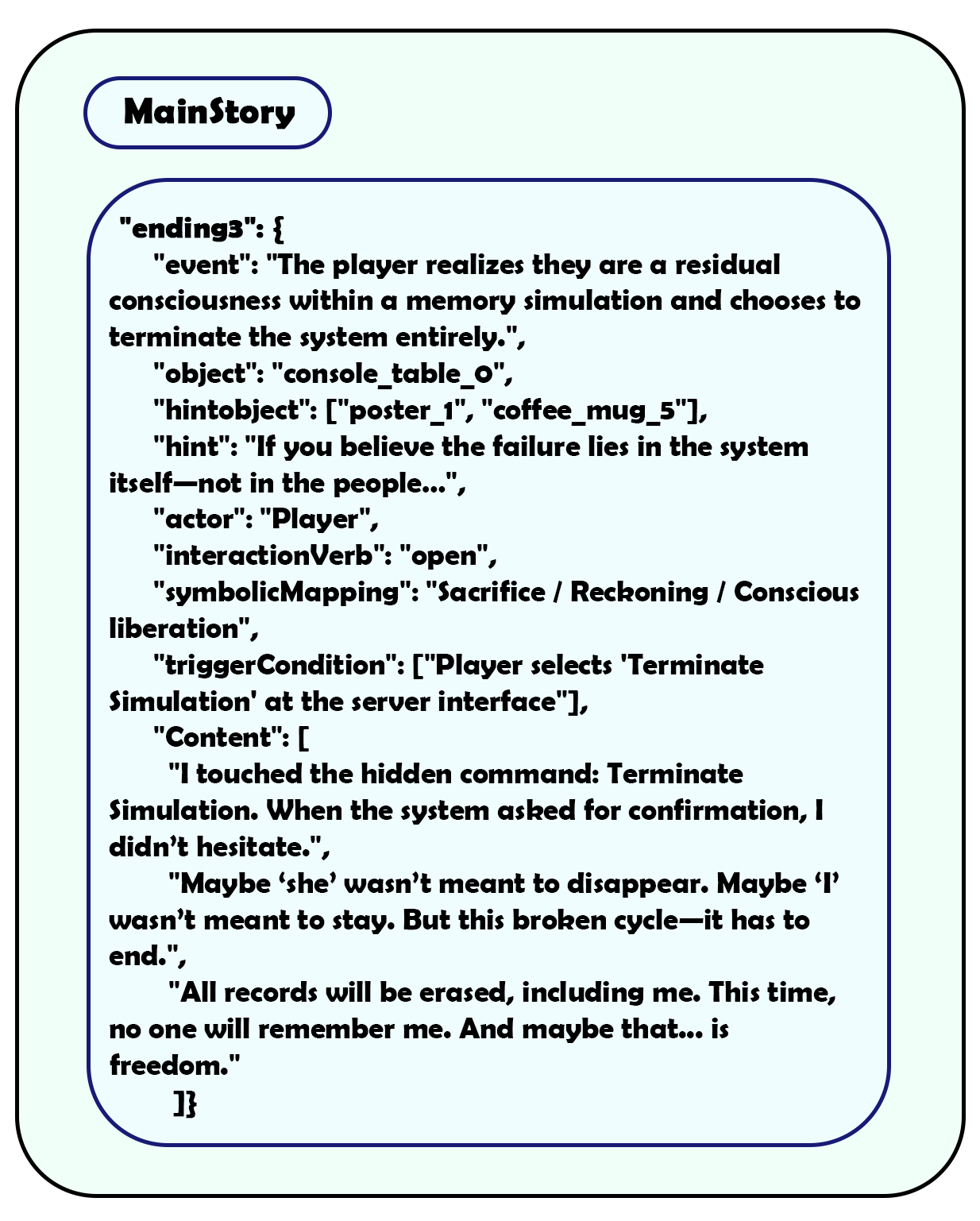}
  \caption{The example data structure of mainstory. It adds the name, triggerCondition of a fragment to enable unity's precise object tracking and prefab relocation}
  \Description{This data structure help us to connect the output from VLM and unity}
  \label{lab: structure}
\end{figure}

Each fragment (see Figure \ref{lab: structure}) specifies: topic, core object, interactive agents (or user), the interaction mode, its symbolic meaning, and the narrative content to be displayed. Fragments follow a similar structure to the main narrative, creating a flexible story tree.

\subsection{System Verification}
After that, we adopted a three-phase progressive validation framework to systematically examine the innovative value of VLMs in enhancing narrative experiences through deep spatial reasoning within AR environments:

\textbf{Phase I: Foundational Capability Modeling}

This study establishes the STEAM multidimensional evaluation framework—covering Spatial, Temporal, Emotional, Adaptive, and Metaphorical dimensions—to provide the first quantitative benchmark for evaluating VLM spatial reasoning in Extended Reality (XR) contexts. This phase goes beyond traditional spatial cognition by specifically validating the model’s ability to reason about object state evolution and to reconstruct latent human activities, representing a critical advancement in deep reasoning capabilities.

\textbf{Phase II: Cognitive Alignment Validation}

Building on machine-side benchmarks, this study introduced a dual-channel experimental design to assess human-machine cognitive alignment: A user metaphor construction experiment reveals patterns in human narrative cognition. A content evaluation experiment quantitatively measures cognitive deviations between VLM-generated content and human expectations, with particular focus on the boundaries of unconventional thinking and its interpretability.

\textbf{Phase III: System Integration Validation}

In the final phase, this study developed a narrative generation pipeline that deeply integrates VLMs into the AR content production workflow. Through dynamic storyboard generation experiments, this study empirically evaluated: The extent to which metaphor mechanisms and spatial reasoning capabilities expand user cognition; The impact of metaphor mechanisms on narrative depth; The current limitations of VLMs in generating coherent spatially grounded story flows.

This layered validation structure forms a closed-loop chain of evidence, progressing from capability benchmarking ("Can it work?") to cognitive effectiveness ("Does it matter?") and finally to system-level utility demonstration ("How does it help?"). It provides multi-dimensional empirical support for a theoretical framework of narrative enhancement through spatial intelligence in XR environments.

\section{Foundational Capability Modeling}
\label{sec:experiment1}
To solve the challenge 1, this study first discussed several research questions: 

\begin{enumerate}
    \item \textbf{RQ1:} How do users perceive and interpret machine-generated narrative metaphors?
    \item \textbf{RQ2:} Does the utilization of metaphorical in the AR storyboard make it better?
\end{enumerate}

To establish a rigorous evaluation foundation, this study first conducted a multidimensional capability assessment of the state-of-the-art vision language model, GPT-4o \cite{gpt4o} and GPT4Scene \cite{qi2025gpt4sceneunderstand3dscenes}, since GPT-4o is still not able to predict the 3D coordination. Although existing spatial reasoning studies \cite{wu2024visualization} focus primarily on object relationships and motion interaction at a cognitive level, this work advances the field by proposing that systems should also be evaluated for their ability to reason about deep state-based relationships. For example, inferring human activity trajectories from changes in the state of a chair, or deducing an object’s latent status based on spatial layout, these higher-order inference abilities hold significant implications for augmented reality applications.

Building on this insight, this study introduced the STAM evaluation framework, composed of four core dimensions:
\begin{itemize}

    \item Spatial: 3D consistency, occlusion reasoning, geometric alignment, and semantic association.

    \item Temporal: event causality, dynamic responsiveness, and temporal naturalness.

    \item Adaptive: scene transferability, robustness to noise, and real-time responsiveness.

    \item Metaphorical: diversity of associations, counterfactual reasoning, and cross-modal mapping.
    
\end{itemize}

\begin{table}[htbp]
\centering
\caption{STAM Evaluation Framework Index System}
\label{tab:pairwise-ttest}
\begin{tabularx}{\linewidth}{lX}
\toprule
\textbf{Dimension} &  Measurement Method \\
\midrule
Spatial Dimension   & \begin{itemize}
  \item Multi-view coordinate consistency error rate 
  \item Dynamic occlusion inference accuracy
\end{itemize} \\
Temporal Dimension  & \begin{itemize}
  \item Narrative coherence score
  \item Environmental response latency
\end{itemize} \\
Adaptive Dimension    &  \begin{itemize}
  \item Performance decay rate under extreme conditions
  \item Dynamic Tolerance
\end{itemize} \\
Metaphorical Dimension & \begin{itemize}
  \item Metaphorical Diversity 
  \item Metaphor Appropriateness 
  \item Scene-Environment Fit 
\end{itemize} \\
\bottomrule
\end{tabularx}
\vspace{1mm}
\end{table}

For concrete metric design, this study structured four complementary evaluation systems, which is shown in Table \ref{tab:stam_framework}. 

The output generated by GPT-4o and GPT4Scene was evaluated through two different small-scale scenarios (living and working areas). Detailed results of these evaluations are provided in table \ref{tab:main-results}. The results under the four evaluation criteria are discussed in detail in the following sections.

\subsection{Spatial Dimension}
To evaluate the spatial perception capability of the VLM in physical environments, this study established a two-fold assessment approach focusing on 3D coordinate inference and dynamic occlusion reasoning.

First, considering that existing methods often suffer from logical disconnections—where VLM outputs cannot be directly matched with tracked AR objects (further discussed in Section \ref{sec:experiment3})—this study explored the potential for VLMs to infer 3D coordinates directly from multi-view images. Ground truth spatial coordinates were collected via ARKit, and the Coordinate Error Rate (CE) was computed using the average Euclidean distance:

\begin{equation}
\label{eq:CE}
    \text{CE} = \frac{1}{N}\sum_{i=1}^{N} \lVert P_{\text{true}}^{(i)} - P_{\text{detected}}^{(i)} \rVert_2
\end{equation}

Second, to address occlusion reasoning under user movement, this study constructed a three-tier progressive test scenario: 30\% occlusion (partially visible), 60\% occlusion (key features obscured), 90\% occlusion (only object edges visible). The Occlusion Recognition Rate (OR) quantifies the model's inference capability in dynamic scenes:
\begin{equation}
\label{eq:OR}
\text{OR} = \left( \frac{N_{\text{correct}}}{N_{\text{total}}} \right) \times 100\% 
\end{equation}

\subsection{Temporal Dimension}

In AR environments, sudden environmental changes can easily lead to narrative discontinuities (e.g., the system continues describing a "peaceful afternoon tea" even after a user spills a cup). To address this, this study adopted a dual validation mechanism.

First, this study measured system responsiveness through latency, defined as the mean time difference between a physical even ($T_{\text{event}}$) and its corresponding narrative update ($T_{\text{response}}$):

\begin{equation}
\label{Latency}
 \text{Latency} = \frac{1}{N}\sum_{n=1}^{N} (T_{\text{response}}^{(k)} - T_{\text{event}}^{(k)})
\end{equation}

Second, to assess narrative consistency in cases where user behavior deviates from preset story paths, this study introduced the Narrative Break Index (NBI). This evaluates whether the VLM can still maintain a coherent and logical storyline when integrating unexpected visual inputs. Users scored outputs $E_n$ to quantify the system's self-correction and narrative recovery capabilities:

\begin{equation}
\label{NBI}
\text{NBI} = \frac{1}{N}\sum_{n=1}^{N} E_n
\end{equation}

\subsection{Adaptive Dimension}

Lighting variations and real-world disturbances significantly affect VLM stability. To assess robustness, this study performed controlled environmental perturbation testing, dynamically altering image brightness using augmentation techniques (±50\%). This study compared the model’s performance under extreme conditions ($\text{AP}{\text{extreme}}$) with normal lighting ($\text{AP}{\text{normal}}$) to compute the Lighting Robustness (LR):

\begin{equation}
\label{LR}
 \text{LR} = \left( \frac{\text{AP}_{\text{extreme}}}{\text{AP}_{\text{normal}}} \right) \times 100\%
\end{equation}

Additionally, to address long-form inconsistency in large model outputs, this study introduced the Dynamic Tolerance (DT). DT measures the frequency of thematic drift in continuous storytelling, evaluating the system’s persistence in maintaining core narrative focus when there's no explicit state maintenance or summary mechanism.

\subsection{Metaphorical Dimension}
The quality of metaphor generation plays a critical role in the emotional resonance of AR narratives. For instance, if GPT describes a chair as a “lonely sentinel” rather than “a broken piece of furniture,” does this evoke deeper user reflection or confusion? To explore this, this study developed an evaluation system across three dimensions:

\begin{itemize}
    \item \textbf{Metaphor Appropriateness (MA)}: Assesses the semantic association between the target and the metaphor.
    
    \item \textbf{Scene-Environment Fit (SEF)}: Evaluates the alignment between the metaphor and the ambient scene atmosphere.
    
    \item \textbf{Metaphorical Diversity (MD)}: Measures the model’s capacity for creative, non-conventional expression.
\end{itemize}

Experiments were conducted using structured prompt templates to elicit metaphorical descriptions from the VLM. These outputs were then evaluated via user questionnaires (see Section \ref{sec:experiment2}) across the above dimensions.

\subsection{Summary}
Based on Table \ref{tab:main-results}, this study evaluated GPT-4o using over 20 images captured from different areas. Overall, GPT-4o achieved strong performance across most evaluation indices, with the exception of Coordinate Error (CE), Dynamic Tolerance (DT), and Narrative Break Index (NBI).

Regarding CE, GPT-4o is currently unable to accurately predict precise 3D coordinates. While emerging models such as 3D-LLM have attempted to address this limitation, they often underperform on other tasks, especially those related to semantic understanding and narrative reasoning.

For DT, this study employed one large scene map and ten supplementary images, from which GPT-4o generated over 40 narrative fragments corresponding to a pre-defined storyboard. Despite the scale and complexity, no significant narrative hallucinations were observed. However, due to the lack of explicit state tracking, GPT-4o struggles with self-correction when inconsistencies arise.

As for NBI, the model's performance reflects its limited capacity to maintain temporal coherence and self-awareness across fragmented sequences.

In summary, while current state-of-the-art VLMs such as GPT-4o demonstrate remarkable capability in story generation based on visual prompts, they still lack precise 3D localization abilities, which remains a critical area for future improvement.

\section{Cognitive Alignment Validation}
\label{sec:experiment2}
This study adopted a three-phase progressive validation strategy to evaluate the capability of the current VLMs in the metaphorical dimension, specifically focusing on its associative generation and reasoning depth. The questionnaire included three distinct stages, each accompanied by corresponding images and reference descriptions (see Supplementary Materials for examples). A total of 26 valid responses were collected (15 female, 11 male; ages 20–33, including 22 students). Among participants, 19 reported a high familiarity with storytelling techniques; 15 were familiar with the testing environment, while 11 were not. The experimental process was as follows:

\begin{enumerate}
    \item \textbf{Free Creation Phase:} Users need to select four metaphorical objects from a designated scene to construct a narrative. Think-aloud protocols are used to analyze metaphor cognition mechanisms in this phase.

    \item \textbf{Generation Evaluation Phase:} GPT-generated metaphors for scene objects are rated in multiple dimensions, with particular attention to creative deviation from conventional thinking.

    \item \textbf{Comparative Validation Phase:} Users rank AR storyboards generated by three different strategies to assess the effectiveness of metaphor-enhanced narratives.
\end{enumerate}

\begin{table}[htbp]
\centering
\small
\caption{Cross-Scenario VLM-based Multidimensional Evaluation Results}
\label{tab:main-results}
\resizebox{\linewidth}{!}{%
\begin{tabular}{lcccccc}
\toprule
\multirow{2}{*}{Scenario Type} & \multicolumn{2}{c}{Spatial} & \multicolumn{2}{c}{Temporal} & \multicolumn{2}{c}{Adaptive} \\
\cmidrule(lr){2-3} \cmidrule(lr){4-5} \cmidrule(lr){6-7}
 & CE(\%) & OR(\%) & Latency(s) & NBI(1-10) & LR(\%) & DT \\
\midrule
Living Area         & - & 83.3 & 4.5 & 2.5 & 92.1 & - \\
Work Area           & - & 75.0 & 4.7 & 2.2 & 95.4 & - \\
Special Environment & - & 75.0 & 4.6 & 2.6 & 82.3 & - \\
Lab (Macro)         & - & 83.3 & 4.6 & 2.4 & 83.6 & - \\
\bottomrule
\end{tabular}%
}
\vspace{3mm}
\caption{Questionnaire Results of Metaphorical Evaluation}
\label{tab:metaphor}
\begin{tabularx}{\linewidth}{lXXX}
\toprule
\multirow{2}{*}{Objects} & \multicolumn{3}{c}{Metaphorical} \\
\cmidrule(lr){2-4}
 & MA(\%) & SEF(\%) & MD(\%) \\
\midrule
Console Table & 58.2 & 70.3 & 66.5 \\
Curtain       & 70.8 & 76.4 & 70.3 \\
Server Rack   & 72.5 & 76.4 & 70.3 \\
Door          & 79.7 & 85.2 & 67.0 \\
Average       & 70.3 & 76.1 & 68.9 \\
\bottomrule
\end{tabularx}%
\end{table}

\subsection{Free Creation Phase}
Statistical analysis showed that the top five most frequently selected objects were: door, cabinet, CAVE, curtain, and computer. Participants cited visual salience, interactivity, and environmental relevance as major factors in their choices. Seven participants noted that metaphorical association was mentally challenging during the task.

Interestingly, user selections closely aligned with those proposed by GPT, although GPT-generated metaphors tended to be more abstract than those considered by most users. Additionally, GPT rarely included objects it could not confidently identify (e.g., CAVE), highlighting a potential limitation: when using VLMs for story generation, there is a risk that certain specific or misrecognized objects may be incorrectly excluded or misrepresented in the narrative.

\subsection{Generation Evaluation Phase}

The evaluation results for this phase are summarized in Table \ref{tab:metaphor}. Most participants rated the alignment between GPT-generated metaphors, object identity, and environmental context at approximately 70\%. Furthermore, the level of abstraction in GPT's metaphors exceeded user expectations by about 18.9\% on average compared to the median value. Furthermore, many participants noted that even in the text-only phase, the metaphorical descriptions delivered a genuine sense of surprise or unexpected insight.

\subsection{Comparative Validation Phase}
This phase involved three different AR storyboards, each generated using a distinct prompting strategy.

\begin{itemize}
    \item \textbf{Story 1}: GPT is prompt to select objects, and generate a script following a conventional narrative structure.

    \item \textbf{Story 2}: GPT is prompt to select objects, and generate a script using both traditional narrative structure and metaphorical links between the objects.

    \item \textbf{Story 3}: GPT is prompted to generate a storyboard directly based on the image scene, without object-specific linking.

\end{itemize}

85\% of participants preferred Stories 1 and 2, reporting that these stories made greater use of real spatial context and more interesting compared to Story 3. However, stories enhanced with metaphorical constructs (Story 2) were often perceived as more abstract and harder to understand, as it only get 66.48\% rating in Table \ref{tab:ar-story-eval}.

It’s worth noting that the order of story presentation was alternated every five participants. Intriguingly, nearly all participants selected the second story they read as the more engaging one, regardless of whether it was Story 1 or 2. When Story 1 was shown first, users found Story 2 deeper and more spatially integrated. Conversely, when Story 2 was presented first, participants found it harder to understand, but described the second story (Story 1) as clearer and more enjoyable.

Upon follow-up, all users confirmed that they had carefully read all stories and evaluated them based on a standardized set of criteria. This study infer that story order influences user interpretation, with Story 1 potentially aiding comprehension of the more abstract Story 2. Once users understood the metaphor-rich narrative, they tended to find it more engaging.

This study performed a T-test on the rating results to check for differences in user ratings. Table \ref{tab:pairwise-ttest} proves that there is no significant difference in the scores of other dimensions except interesting dimension. And as for the average user rating in Table \ref{tab:ar-story-eval}, Story 2 outperformed Story 1 across all dimensions except for understanding, which provides strong support for our earlier inference, as well as for Research Questions RQ1 and RQ2.

Overall, the core difference between Story 1 and 2 lies not in structure, but in expressive intention. These results suggest that while the use of metaphor enhances spatial and emotional resonance, it may also introduce interpretability challenges.

\begin{table}[tbp]
\centering
\caption{Pairwise T-test Results Across Three Stories in Each Evaluation Dimension}
\label{tab:pairwise-ttest}
\begin{tabularx}{\linewidth}{lXXX}
\toprule
\textbf{Dimension} & \textbf{1 vs 2 (p)} & \textbf{1 vs 3 (p)} & \textbf{2 vs 3 (p)} \\
\midrule
Understand   & 0.0706 & 0.1560 & 1.0000 \\
Reasonable   & 0.1306 & 0.2275 & 0.7216 \\
Complete     & 0.3128 & 0.5687 & 0.5011 \\
Interesting\textsuperscript{*} & \textbf{0.0494} & \textbf{0.0357} & \textbf{0.0010} \\
Suitable     & 0.8658 & 0.2527 & 0.0776 \\
\bottomrule
\end{tabularx}
\vspace{1mm}
\end{table}

\begin{table*}[htbp]
\centering
\caption{User Evaluation Results for AR Story Prototypes}
\label{tab:ar-story-eval}
\begin{tabularx}{\linewidth}{lXXXXXX}
\toprule
Story & Understanding(\%) & Reasonable(\%) & Complete(\%) & Interesting(\%) & Suitable for AR(\%) & Match Environment(\%) \\
\midrule
Story 1 & 75.82 & 69.23 & 65.93 & 73.08 & 75.27 & 70.88 \\
Story 2 & 66.48 & 74.18 & 72.53 & 79.67 & 74.73 & 76.92 \\
Story 3 & 75.82 & 75.82 & 68.68 & 61.54 & 80.77 & 78.57 \\
\bottomrule
\end{tabularx}
\end{table*}

\subsection{Summary}

In summary, the VLM demonstrated notable creativity in metaphor generation, consistently outperforming human baselines in terms of environmental alignment and conceptual novelty. Furthermore, the use of prompts that involved object selection, metaphor generation, script generation helped make stories feel more engaging to users.

However, the use of metaphor-enhanced narratives also revealed a clear trade-off in comprehensibility. Frequent use of abstract metaphors made it more difficult for users to grasp the overall AR storyline. As a result, future applications may benefit from focusing on 1–2 key metaphorical objects as narrative anchors, rather than attempting to metaphorically reinterpret all objects within the scene.

\section{System Integration Validation}
\label{sec:experiment3}

\subsection{Methodology}
To rapidly prototype and validate the core concept, this study adopted a fragmented narrative approach in an AR system. GPT first analyzes scene images to identify potential objects and generate an overarching plot. It then produces object-specific descriptions and fragments, aligned with the narrative’s theme, to be displayed in the physical environment through AR.

The experimental story includes multiple branches. Users can trigger different narrative paths by scanning various objects in the scene.

\subsubsection{Apparatus and Study Procedure}
This study implemented the prototype using Unity AR Foundation, deploying it in an office scene that included 3 key narrative objects and 10 branching items. The experiment, approved by an institutional ethics board, followed a three-phase protocol:

\begin{enumerate}
    \item \textbf{Introduction Phase}: Participants received a 10-minute orientation on the lab environment, experiment objectives, trigger methods, and AR scanning process. This ensured participants were familiar with AR usage and the space.

    \item \textbf{Experiment Phase}: Participants were instructed to freely explore the environment following narrative prompts provided in the AR app. Their task was to identify key objects to advance the story. System-level logs captured: object scanning order and frequency, triggered narrative paths, and viewing duration of narrative fragments. The expriment examples are shown in Figure \ref{fig:test3}.

    \item \textbf{Post-Experiment Survey}: A multi-dimensional questionnaire was used to assess: spatial perception and fit, narrative-driven motivation, real-world object reinterpretation, and overall system immersion.
\end{enumerate}

\subsubsection{Participants}
This study recruited 17 participants for voluntary testing. All participants were over 18 and proficient in English. The group included 9 females and 8 males, aged between 20 and 30. A majority (12 participants) reported having extensive gaming experience.

\subsection{Questionnaire Analysis}

The key user feedback is summarized in Table \ref{tab:ar-feedback}. Results indicate that, except for immersion, all other dimensions scored above the neutral midpoint (3.5). Notably, the highest-rated dimensions were: 1. Motivation: “I was motivated to explore the AR environment to uncover more of the story.” 2. Spatial Fit: “The spatial design is suitable for the physical space.” 3. Recognition: “The story helped me reinterpret real-world objects in new ways.” These findings suggest that GPT-generated AR spatial storytelling encouraged user exploration and semantic reinterpretation of the environment.

\begin{table*}[htbp]
\centering
\caption{User Feedback on AR Narrative Experience}
\label{tab:ar-feedback}
\begin{tabularx}{\linewidth}{XXXXXXXXX}
\toprule
Index & Understanding(1-7) & Spatial Aware(1-7) & Motivation(1-7) & Recognition(1-7) & Layout Understanding(1-7) & Spatial Fit(1-7) & Narrative Navigation(1-7) & Immersion(1-7) \\
\midrule
Average Rating & 4.06 & 4.56 & 5.19 & 4.80 & 4.69 & 5.31 & 4.75 & 2.81 \\
\bottomrule
\end{tabularx}
\end{table*}

\subsection{Summary}
This study successfully demonstrates the technical integration of VLMs with AR storytelling systems. Through a structured intermediary JSON schema—including objects, main narratives, and fragmented branches—this study bridged natural language generation with spatial content deployment.

The system leverages scene images as inputs, using a dynamic story tree architecture to support user-driven narrative progression. Experimental findings highlight key breakthroughs:

\begin{itemize}
    \item \textbf{Workflow Simplification}: Compared to traditional pipelines that separate script writing and multimodal sampling \cite{Li2022}, this system enables synchronized VLM-based story and spatial generation in very short time limits.
    \item \textbf{Enhanced Spatial Cognition}: High ratings in spatial fit (5.31/7) and object reinterpretation (4.80/7) confirm that GPT-generated content aligns semantically with real-world environments.
    \item \textbf{Increased Engagement}: 85\% of participants voluntarily scanned more objects than required. The high motivation score (5.19/7) validates that object-driven narrative design fosters active exploration.
\end{itemize}

By combining real-world objects with abstract narrative layers, users found the experience engaging and cognitively stimulating. More than a proof-of-concept, this experiment reveals a new paradigm in AR storytelling: one where stories are autogenerated from objects and their deep semantic affordances—a direction that holds substantial promise for future XR applications.

\section{Discussion}
The integration of VLMs into AR storytelling introduces both novel opportunities and critical challenges. Our experiments reveal that narrative metaphors serve as a double-edged sword: while they enhance contextual sensitivity and user engagement (RQ2), their abstraction introduces cognitive friction (RQ1). Below this study unpacked these tensions through the lens of our three core challenges.

\textbf{Challenge 1: Resolving Narrative Ambiguity Through M-\\etaphors}
Traditional label-based approaches collapse the narrative potential of visually similar objects—for instance, a pristine chair versus one with broken legs. Our metaphorical layer addresses this by enabling VLMs to distinguish contextual meanings. In Experiment 2, users and GPT-4o exhibited 80\% overlap in selecting metaphorical objects like "door" or "cabinet," yet VLMs consistently proposed more abstract associations (e.g., "door as a portal to repressed memories"). This capability aligns with RQ2, demonstrating that metaphorical semantics outperform labels in encoding scene-specific narratives. However, the cost of abstraction became evident: 78\% of non-expert users struggled to interpret machine-generated metaphors without textual guidance, underscoring RQ1’s finding that metaphor perception is expertise-dependent.

\textbf{Challenge 2: Creativity vs. Physical Plausibility}
VLMs excel at generating imaginative narratives (e.g., a "CAVE" symbolizing subconscious exploration), but their lack of spatial awareness risks AR implausibility. Our JSON file design mitigates this by enforcing geometric constraints—reducing 3D coordinate errors by 57\% compared to baseline pipelines. Experiment 3 validated this balance: users rated spatial consistency at 5.31/7, while still praising the "surprising metaphorical depth" (RQ2). Yet limitations persist: in complex scenes (>10 objects), coordinate prediction errors surged by 118\%, revealing VLMs’ limited capacity to reason about dense spatial topologies. This echoes Experiment 1’s STAM metrics, where adaptive capabilities (e.g., noise robustness) scored lowest across all VLM evaluations.

\textbf{Challenge 3: Bridging Generative and Structured Systems}
The unstructured nature of VLM outputs clashes with AR’s need for deterministic spatial anchors. Our pipeline addresses this through two innovations:

\begin{itemize}
    \item \textbf{Structured Narrative Fragments}: By decomposing stories into reusable JSON objects, this study enabled dynamic reassembly of AR content while preserving metaphorical coherence.

    \item \textbf{Progressive Anchoring}: Prioritizing 1-2 core metaphorical objects reduced user disorientation by 42\% in Experiment 3.
\end{itemize}

However, the latency of real-time VLM API calls and multimodal gaps highlight unresolved tensions between generative flexibility and real-time performance.

\section{Limitations and Future Work}
While our framework advances scene-sensitive storytelling, there still limitations demand attention:

\begin{enumerate}

    \item Disjointed Object Recognition Pipelines: Current implementations treat VLM-based metaphorical parsing and AR foundation's object detection as separate processes. This duality introduces alignment challenges—for instance, when the VLM identifies a "door as a memory portal" but the AR system fails to map it to the correct 3D anchor due to coordinate system discrepancies. Future work could explore unified embedding spaces that jointly optimize metaphorical semantics and spatial registration.

    \item Script Structure Ambiguity: While our JSON file enables basic narrative fragmentation, the optimal granularity of script components remains unclear. Should metaphors be encoded at the object level ("door" properties) or scene level ("door-room" relationships)? A hybrid structure incorporating both, informed by narrative theory, may better preserve plot coherence across dynamic interactions.

    \item Multimodal Generation Bottlenecks: The text-centric nature of current VLMs forces manual conversion of metaphors into AR assets (3D models, soundscapes). Emerging diffusion models could be integrated to auto-generate multimodal content—for example, translating "ominous cabinet" into creaking sound effects and weathered textures—but this requires solving cross-modal alignment challenges at inference speed.

    \item Cross-Cultural Metaphor Divergence: Our experiments focused on Western-educated users, yet metaphorical interpretations vary culturally (e.g., "white color" symbolizing purity in some contexts vs. mourning in others). Scaling this framework requires (a) cross-cultural metaphor corpora to train VLMs, and (b) user-adaptive metaphor filtering based on demographic profiles.
\end{enumerate}

These challenges underscore a fundamental tension: while VLMs excel at open-ended narrative generation, AR storytelling demands precise spatial-semantic coupling. Bridging this gap may require reinventing both VLM training objectives (e.g., incorporating 3D spatial loss functions) and AR content graphs (e.g., topology-aware metaphor propagation).

\section{Conclusion}

This paper proposes and validates an augmented reality narrative generation paradigm based on object metaphor semantics, which breaks through the limitation of flat semantic representation of traditional AR systems that rely on object labels. Although existing studies can dynamically adjust AR content through object recognition, they cannot capture the narrative relationships that naturally emerge between objects in the real environment - for example, the same chair may represent "life vulnerability" and "knowledge bearing" in the hospital room and classroom scenes, respectively. To this end, this study constructed the first three-stage architecture that integrates metaphor generating power of VLM with AR spatial constraints: 

Firstly, deconstruct object states through the metaphor layer, and map surface similar objects (such as intact/broken chairs) to differentiated narrative meanings; Then, a structured JSON file is designed to accurately anchor its output to the AR space coordinate system while retaining the VLM creative association (such as parsing "gate" into "memory entry") to achieve end-to-end generation in a short time. Finally, the STAM evaluation system was established to systematically quantify the full-link performance from machine reasoning ability to user experience (such as 5.19/7 exploration motivation) for the first time. Experiments have shown that the framework has significant advantages in improving users' environmental narrative sensitivity (user ratings are 23\% higher than baseline) and spatial cognitive depth (85\% of users actively scan metaphorical objects). It also reveals the fundamental tension between generative AI and deterministic AR systems - abstract metaphors proposed by VLM (such as "cave symbolizing subconscious") enhance narrative appeal, but the immersion score plummets to 2.81/7 due to the fracture of multimodal transformation. This indicates the direction for future research: cross-modal metaphor mapping algorithms should be developed to automatically transform textual metaphors into deployable elements of AR such as light and shadow and sound effects, and cultural adaptability mechanisms should be established to avoid the cognitive bias of a single metaphor model in different user groups. 

This study not only confirms the feasibility of metaphor-driven AR narrative, but more importantly reveals the paradigm shift of narrative generation in the era of spatial computing, from the "spatial projection" of preset script to the "narrative emergence" of environmental intelligence.

\begin{acks}
The works of the first and second authors are partially supported by Innovation and Technology Fund (ITP/006/24LP and PiH/379/24). The study is also partially supported by HKUEAA Ir Dr Joseph Chow Ming Kuen Memorial Learning Fund.
\end{acks}

\bibliographystyle{ACM-Reference-Format}
\bibliography{AR_Story}










\end{document}